# Demand Forecasting in the Presence of Systematic Events: Cases in Capturing Sales Promotions

Mahdi Abolghasemi[1], Ali Eshragh[1], Jason Hurley[2], Behnam Fahimnia[2]

[1]School of Mathematical and Physical Sciences, The University of Newcastle, NSW, Australia.

[2]Institute of Transport and Logistics Studies, The University of Sydney, NSW, Australia

## Abstract

Reliable demand forecasts are critical for effective supply chain management. Several endogenous and exogenous variables can influence the dynamics of demand, and hence a single statistical model that only consists of historical sales data is often insufficient to produce accurate forecasts. In practice, the forecasts generated by baseline statistical models are often judgmentally adjusted by forecasters to incorporate factors and information that are not incorporated in the baseline models. There are however systematic events whose effect can be effectively quantified and modeled to help minimize human intervention in adjusting the baseline forecasts. In this paper, we develop and test a novel regime-switching approach to quantify systematic information/events and objectively incorporate them into the baseline statistical model. Our simple yet practical and effective model can help limit forecast adjustments to only focus on the impact of less systematic events such as sudden climate change or dynamic market activities. The proposed model and approach is validated empirically using sales and promotional data from two Australian companies. Discussions focus on thorough analysis of the forecasting and benchmarking results. Our analysis indicates that the proposed model can successfully improve the forecast accuracy when compared to the current industry practice which heavily relies on human judgment to factor in all types of information/events.

# 1 Introduction

Demand forecasts[1] are critical pieces of information in supply chain management because numerous decisions – such as sourcing, production planning, logistics, inventory management and retail decisions – heavily rely on forecasts. In particular, demand forecasting is a key ingredient in sales and operations planning (S&OP) which is responsible for continuous alignment between demand plans and supply plans (Fildes, Goodwin, & Önkal, 2018). Therefore, improving the accuracy of product demand forecasts can directly result in better operational efficiency, customer satisfaction, and financial savings throughout the entire supply chain (Kremer, Siemsen, & Thomas, 2015; Trapero, Kourentzes, & Fildes, 2015).

Having historical demand information is beneficial for generating accurate forecasts, albeit is often not solely sufficient to forecast to a desired degree of precision (Hyndman & Athanasopoulos, 2014). This is because many statistical forecasting models premised on historical data lack the ability to explicitly capture the contextual information[2] and/or dynamically update as more recent information becomes available (Lawrence, Goodwin, O'Connor, & Önkal, 2006). Special events such as marketing campaigns, holidays and sales promotions are examples of valuable information often not incorporated into univariate statistical forecasting models. Particularly, retailer sales promotions have been shown to significantly influence consumer behavior and market demand (Trapero et al., 2015; Trapero, Pedregal, Fildes, & Kourentzes, 2013). Since events such as promotions can lead to non-stationary time series, single static time series forecasting methods may not be the most suitable. Hence in practice, the output of such methods are merely used as baseline forecasts which are subject to judgmental adjustment by sales forecasters (Fildes, Goodwin, Lawrence, & Nikolopoulos, 2009).

Using expert judgment in complement to statistically analyzing large amounts of data has been shown to be beneficial for improving forecast accuracy (Alvarado-Valencia, Barrero, Önkal, & Dennerlein, 2017). Moreover, evidence suggests that the human input to forecasts can be improved by providing a systematic approach to structure the information utilized when imposing judgment to make adjustments (Franses & Legerstee, 2013). For instance, a structured-analogies method is shown to lead to more accurate forecasts than when produced with unaided judgment (Green & Armstrong, 2007).

---

[1] Demand forecasting and sales forecasting are used interchangeably throughout this paper.

[2] *Contextual information* is referred to non-time series information that is highly relevant to interpreting, explaining and anticipating time series behavior.

Yet no optimal procedure exists for structuring the critical information that forecasters must consider and there is no definitive answer regarding how to most effectively integrate human judgment with forecasting models (Baecke, De Baets, & Vanderheyden, 2017).

In this paper, we aim to contribute to this area by developing and validating a forecasting model that can capture the effects of quantifiable systematic events which would otherwise require judgmental adjustments. More precisely, we develop a time series regression model that incorporates some of the most significant factors/information considered by forecasters when adjusting baseline forecasts. The model takes into account the dynamics of historical base sales and the additional influential factors. Our observations in the Fast-Moving Consumer Goods (FMCG) industry have revealed that retailer sales promotions are the main reason for forecast adjustments. This is also supported by the academic literature (e.g., Fildes & Goodwin, 2007; Fildes et al., 2009). Therefore, we utilize the proposed model to deal with sales promotions, a good exemplar of systematic events. The model is able to systematically define various states of demand uplift by analyzing historical sales data and different combinations of promotions. The obtained demand states can then be embedded into the model. Although we explore the application of our model to deal with sales promotions, the underlying algorithm could be adapted to consider other systematic events such as holidays and seasonality trends.

The proposed model is validated in two case studies and its capability to improve forecast accuracy compared to current industry practice is demonstrated. The sales data was obtained from two giant FMCG companies in Australia. Even though both businesses operate within the food and beverage industry, their product and promotion characteristics substantially differ in terms of product perishability and usage, as well as promotion frequency and magnitude of demand uplift, which to some extent helps the generalizability of our model application and study findings. This model can potentially aid sales forecasters and demand planners by reducing the complexity of the forecasting task and ease the cognitive load associated with processing vast amounts of unmodeled information (Lawrence et al., 2006).

The remainder of this paper is structured as follows. Section 2 reviews the literature related to demand forecasting in a supply chain context, including the main quantitative and judgmental approaches as well as how they handle the impact of sales promotions. Section 3 describes the methodology employed in this study and explains the structure of the proposed time series regression model. Section 4 focuses on validating the model and methodology using two empirical case studies. Concluding remarks are presented in Section 5 including the study limitations and future research directions.

## 2   Related Literature

A myriad of forecasting methods and models have been developed with the common goal of improving accuracy. In the following sub-sections, we review some of the quantitative (statistical and analytical) and judgmental forecasting approaches that are particularly concerned with addressing the impact of sales promotions.

### 2.1   Quantitative Forecasting Methods

The evolution of statistical forecasting methods has largely been driven by advancements in computational power, software and information system technologies such as enterprise resource planning systems, electronic data interchange, and point of sale scanning (Sanders & Manrodt, 2003a). Such advancements have enabled vast amounts of data/information to be easily collected, utilized in more sophisticated statistical models, and shared throughout the supply chain. The core of most quantitative approaches to forecasting is extrapolation (Fildes, Nikolopoulos, Crone, & Syntetos, 2008). Extrapolative methods use purely historical data to predict the future. Of the extrapolative methods, exponential smoothing is one of the classical forecasting techniques and is widely practiced in industry (Fildes, 1992; Hyndman & Koehler, 2006). Exponential smoothing is a statistical technique that averages (smooths) time series data, differing from a simple moving average in that it assigns a larger weight to recent observations, and exponentially decreases the weight of observations over time. Several different variations of exponential smoothing exist (e.g., simple, Holt, Pegels, Holt-Winters, and variants of these with damped trends), where each method is suited for different forecast horizons (i.e., short-term or long-range), seasonality types and trends in the time series (i.e., constant, additive or multiplicative) (Taylor, 2003). Furthermore, the Auto-Regressive Integrated Moving Average (ARIMA) model (also known as Box-Jenkins model) along with its numerous variants (e.g., SARIMA, ARIMAX, ARMA-GARCH, ARFIMA) are also widely utilized extrapolative methods as they can account for trends, seasonality, errors and non-stationary aspects of a time series (Nikolopoulos, Syntetos, Boylan, Petropoulos, & Assimakopoulos, 2011).

The second main statistical forecasting approach is causal and multivariate methods, which have been largely developed by the study of econometrics data (Fildes et al., 2008). These methods are forms of regression analysis and assume that there is a cause-and-effect relationship between the dependent variable (i.e., demand) and one or more independent variables (explanatory factors influencing the demand). Causal and multivariate methods are capable of addressing the issue of a non-stationary time series by considering different exogenous variables such as promotions, holidays and special events that can impact customer demand (Fildes et al., 2008; Trapero et al., 2013). While there are many variables and vast amounts of information that can be included in forecasts, it is prudent to keep models as parsimonious as possible while maintaining desired accuracy. This is because multivariate models with a high number of explanatory variables have large data requirements, in addition to being prone to multicollinearity and dimensionality problems (Trapero et al., 2015).

*Computer-intensive methods* are much more contemporary than those discussed above since these analytical methods rely on intelligent software and computational power that was unfathomable just decades ago. Moreover, these methods began to attract attention from the operations research community since the early 2000's as predictive analytics and data mining have rapidly grown in popularity (Olafsson, Li, & Wu, 2008). Some of notable computer-intensive forecasting methods include: Artificial Neural Networks (ANN), Support Vector Machines (SVM) and Decision Trees (DT) (R. Fildes et al., 2008). The ANN are non-linear, semi parametric methods which originated with an analogy to the biological nervous system but are now applied to a wide range of applications in business and data mining (Zhang, Patuwo, & Hu, 1998). The SVM and DT are also commonly employed methods since they have the advantage of combining individual classification with a regression component (Murthy, 1998; Smola & Schölkopf, 2004).

Despite the rapid evolution of technology and numerous advancements in statistical and analytical forecasting methods, large-scale empirical evidence from three seminal forecasting competitions (Makridakis et al., 1982; Makridakis et al., 1993; Makridakis & Hibon, 2000) consistently finds that "statistically sophisticated or complex methods do not necessarily produce more accurate forecasts than simpler ones" (Makridakis & Hibon, 2000, p. 452). This notion is also supported by Green and Armstrong (2015) who find that complexity of the forecasting method harms accuracy, and that simpler methods reduce the likelihood of errors as well as better aid the understanding of decision-makers. Furthermore, complex forecasting techniques are not frequently utilized in industry due to high costs, lack of internal expertise and resources, as well as other organizational barriers (Trapero et al., 2015). We aim to address this concern in our study by presenting a simple and practical statistical model which captures the impact of systematic events, and yet still allows the forecaster to intervene and judgmentally incorporate the impact of less quantifiable contextual information. Therefore, the model and approach presented in this paper can be used as a more complete baseline forecast which would require fewer judgmental adjustments, if any at all. A substantial amount of research has been conducted over the years pertaining to judgmental forecasting and the integration of human judgment into statistical models. We briefly review this literature in Section 2.2.

## 2.2   The Human Factor in Forecasting

Academic literature has increasingly acknowledged the importance of human judgment in forecasting as they are highly connected (Lawrence et al., 2006). Despite the broadly acknowledged human factor in forecasting, much of the research on forecasting methods that emerged prior to the 1990's advised against the use of judgment in forecasting (e.g., Armstrong, 1986; Hogarth & Makridakis, 1981). However, recent research advocates that statistical methods and human judgment should be integrated so that complementary benefits can be realized to mitigate the inherent weaknesses of each approach (Alvarado-Valencia et al., 2017; Baecke et al., 2017; Blattberg & Hoch, 1990; Fischer & Harvey, 1999; Franses, 2008; Marmier & Cheikhrouhou, 2010).

One important benefit of integrating statistical methods and human judgment relates to their capability to handle different types of information. Lawrence, O'Connor, and Edmundson (2000) classify the information that is useful for forecasting into two classes: (1) historical data, and (2) contextual or domain knowledge. Historical data is simply the time series of historical product sales that has been recorded, and contextual knowledge being any other information relevant to interpreting, explaining and anticipating time series behavior. Examples of contextual information include: changes in promotional plans, competitor activities, market intelligence, sudden climate changes and dynamic influencers (e.g., political, media/press release, natural or manmade disasters).

Contextual information is the primary factor that leads to instances when judgment is superior to statistical models (Webby & O'Connor, 1996). In fact, the major value-add that comes from human input is because forecasters possess contextual information, intimate product knowledge and experience that statistical models do not (Edmundson, Lawrence, & O'Connor, 1988). Statistical methods are well suited to handle vast amounts of historical data, but when the effects of discontinuities (often caused by contextual factors) cannot be estimated from historical data, statistical methods tend to produce forecasts with sub-optimal accuracy (Gardner, 1985). Human judgment can be utilized to overcome this issue and incorporate valuable contextual information by adjusting baseline statistical forecasts, albeit with caution as there are also human factors that can hinder judgment (e.g., personal or social biases, heuristics, cognitive limitations, and system neglect) (Goodwin, 2002; Kremer et al., 2015; Lawrence et al., 2006).

The great efficiency and flexibility as well as demonstrated accuracy improvements that are realized through judgmental forecast adjustments (e.g., Fildes et al., 2009; Moritz, Siemsen, & Kremer, 2014) have inevitably resulted in its widespread use in industry (Sanders & Manrodt, 2003b). Overall, the debate surrounding the practice of judgmental forecast adjustments has now evolved beyond merely whether they should be utilized or abandoned. The question is more how to appropriately use judgment to consistently improve the accuracy of forecasts. A common practical issue is that forecast adjustments are rarely performed systematically (Trapero et al., 2013), with experts often applying their knowledge and experience in an unaided and unstructured form (Green & Armstrong, 2007). This may lead to poor forecasting outcomes when compared to a structured approach. Ideally, a Forecast Support System (FSS) can be utilized to provide a baseline statistical forecast as well as structured guidance and feedback to effectively inform a forecaster (Fildes, Goodwin, & Lawrence, 2006; Goodwin, Fildes, Lawrence, & Stephens, 2011). The use of an FSS helps ease the cognitive burden on the human mind and consequently improve the accuracy of final forecasts (Adya & Lusk, 2016). The model developed and tested in this paper can act as the core of such an FSS. Adding additional features such as adaptive guidance to this base model can help develop a comprehensive and practical FSS.

## 2.3   Promotional Effects

Sales promotions are common phenomena in contemporary retail operations. Evidence suggests that promotions are the leading cause for judgmental adjustments to statistical forecasts (Fildes & Goodwin, 2007; Goodwin, 2002). When a promotion occurs, a price discount is offered to customers for a specified time-period and a variety of additional actions are also taken to increase the prominence of a given product or service. The additional actions taken are associated with the *promotional mechanics*, which may include: type of promotion (e.g., single-buy, buy one get one free, multi-buy), display type (e.g., front of store, end of aisle), advertisement type (e.g., in-store, online, catalogue), and special features to coincide with holidays/events (e.g., Christmas oriented product labelling, free event-oriented gift with purchase).

There is normally an uplift in sales when promotions are offered. The uplift is often associated with purchasing acceleration, increased consumption and/or brand switching (Blattberg & Neslin, 1989). Moreover, consumers commonly stockpile products while they are on promotion (that is more the case for less perishable items) which often leads to lower sales in the following period(s). Different combinations of promotions result in different sales uplift, but the magnitude of the impact is associated with a high degree of uncertainty given the dynamic nature of consumer behavior. Inevitably, such promotional effects complicate the forecasting process.

The impact of sales promotions on demand has been previously explored (see for example Ali, Sayın, Van Woensel, and Fransoo (2009); Nikolopoulos, Litsa, Petropoulos, Bougioukos, and Khammash (2015); Ramanathan (2012); Ramanathan and Muyldermans (2011); Trapero et al. (2013)) yet quantifying the impact of promotions still proves to be problematic for practicing forecasters and academic researchers alike. There are several reasons why human judgment has been used in promotional sales forecasting. First, univariate statistical methods (e.g., exponential smoothing) only consider historical data and therefore do not account for the effects of future sales promotions in forecasts, unless promotions and corresponding effects are very consistent over time (Trapero et al., 2015). Although such methods are well suited for semi-automatically generating forecasts for numerous products, subsequent judgmental adjustment to account for contextual information is required. Second, judgment is particularly useful when little or no historical data is available such as when a new product or promotional campaign is offered (Oliva & Watson, 2009; Seifert, Siemsen, Hadida, & Eisingerich, 2015). Judgment could also be beneficial in situations where large spikes in demand occur (significant promotions), because univariate statistical models disperse the effects of large changes over the entire horizon (Sanders, 1992). This can result in inaccurate parameter estimation for promotional versus non-promotional periods.

Sophisticated causal methods have also been proposed to handle the task of promotional forecasting (Fildes et al., 2008). These models are usually based on multiple regression with exogenous variables corresponding to various types of promotions. As opposed to judgmental forecasting which is practical and require few resources, such methods are highly complex, have demanding data requirements, and are difficult to interpret in terms of distinguishing the impact of individual promotional variables (Blattberg & Neslin, 1989; Trapero et al., 2015). Nevertheless, sophisticated forecasting models that account for the effects of promotions have been developed (Huang, Fildes, & Soopramanien, 2014; Kourentzes & Petropoulos, 2016), in addition to promotional FSSs such as 'SCAN*PRO' (Van Heerde, Leeflang, & Wittink, 2002), 'PromoCastTM' (Cooper, Baron, Levy, Swisher, & Gogos, 1999), and 'CHAN4CAST' (Divakar, Ratchford, & Shankar, 2005). There have also been attempts to innovate promotional modeling techniques by utilizing structural equation models (Ramanathan & Muyldermans, 2010) and dynamic regression involving principal component analysis and transfer functions (Trapero et al., 2015). Despite all those efforts, evidence indicates that lack of resources, expertise, and high costs hinder the widespread implementation of such methods and support systems in practice (Hughes, 2001).

We aim to tackle this issue by introducing an easy-to-implement and practical model that can be used to incorporate the impact of systematic promotions into the statistical models. Promotions are a good example of systematic events, the impact of which could be quantified and incorporated into the statistical models to provide the forecaster with a more solid and accurate baseline forecast. This enables a forecaster to only focus on incorporating dynamic information and less systematic events whose level of impact requires expert opinion and market intelligence. Our aim in this paper is to develop and validate a new model that is 'simple' and yet 'practical' to produce baseline statistical forecasts which may then only require minor judgmental adjustments. The methodology utilized to do so is discussed next.

# 3    Methodology

In the previous section we described some of the key demand forecasting approaches and how sales promotions can dramatically alter the behavior of consumer demand. We have realized that similar scenarios where time series behavior is subject to constant changes have been studied in a macroeconomic context (e.g., Giordani & Kohn, 2008; Giordani, Kohn, & van Dijk, 2007; Hamilton, 1989, 1990; Kim, Piger, & Startz, 2008) and some of techniques and concepts utilized to tackle those situations could be applicable to a demand forecasting context. In particular, Hamilton (1989) developed a novel approach, the so-called Markov switching model, to more accurately capture and predict changes in the regime or state[3] of non-stationary time series. Although, Hamilton initially applied his model to the United States gross national product data, his approach has provided the foundation to forecast the future values of any time series that exhibits a regime-switching behavior.

We posit that in a product demand forecasting context, sales promotions can cause the time series to abruptly enter different states. Therefore, the concept of Hamilton's model can be applied to define demand states by which we can structure systematic promotional information and embed it in a statistical forecasting model. Motivated by Hamilton's regime switching idea, we develop a time series regression model by exploiting the time series data and information related to systematic events to capture sales dynamics in all periods and forecast future demand. To the best of our knowledge, this is the first attempt in demand forecasting literature to use a regime-switching approach to quantify the impact of systematic events.

In this model, a list of potential systematic events along with their possible levels of impact on the demand are first obtained from expert forecasters. Then, the most significant levels of systematic events are identified through statistical analysis – e.g., through Analysis of Variance (ANOVA). For instance, the most significant systematic event in the retail industry is *promotion type* with two possible levels of major and minor (as is the case in the case studies presented in this paper). Next, all possible combinations of the levels of systematic events are constructed where each combination is called a *state*. Demand uplifts in promotional periods are computed by subtracting baseline forecasts from the actual realized sales value in each epoch. In this paper, we use historical baseline forecasts and the corresponding realized sales figures provided by the case companies. Both case companies generate their baseline forecasts using simple exponential smoothing. However, baseline forecasts can be estimated with numerous models such as those discussed in Section 2.1 (e.g., ARIMA, causal and multivariate methods). The procedure to establish various Demand Uplift States (DUS) can be summarized in the DUS algorithm as follows:

*Step 0.* Input demand time series and a list of potential systematic events along with their possible levels.

*Step 1.* Find the most significant systematic events by running ANOVA over the data.

---

[3] The terms 'regime' and 'state' are used interchangeably throughout this paper and are defined as "episodes across which the dynamic behavior of the series is markedly different" (Hamilton 1989, p. 358).

*Step 2.* Construct all possible combinations of the levels of significant systematic events and label them from *1* to *k*, where *k* is the total number of combinations.

*Step 3.* For i=1 to *k* do

> Compute the average demand uplift for the i^th combination.

> end for

*Step 4.* Put each combination with distinct uplifts in demand in one state.

**return** Demand uplift states labelled *1* to *m*, where *m* is the total number of states constructed in Step 4.

Let us consider the case of sales promotions in retail industry as a case example for a significant systematic event. The common practice in the retail industry is for the retailers and suppliers to negotiate and set promotional plans well in advance of their occurrence. For instance, the case companies in this paper lock in their promotional calendars for the following calendar year. In addition to timing of each promotion, the specific promotion types (also referred to as promotional mechanics) are also decided. However, promotional plans may be altered during the year for a variety of reasons. Examples include extending the well-performing promotions, adding extra promotions to induce sales for stock that is near expiration, and changing dates of promotions due to inclement weather. But final changes are normally locked at least four weeks prior to the promotion commencement. Therefore, given that promotional plans are finalized prior to preparing a forecast, the DUS algorithm can be effectively utilized to determine future states. If any change in promotional plans is realized, the algorithm can be easily re-run to accommodate abrupt variations. If a new significant level of the combination of promotions is offered, based on the potential impact on consumer behavior, forecasters can either assign it to one of the current demand uplift states (Step 4 of the DUS algorithm), or define a new state depending on the magnitude of its effect.

Following the DUS algorithm, we introduce our new time series regression model, named Forecasting Systematic Events (FSE) model, formulated in Equation (1).

$$X_t = \alpha_0 + \sum_{i=1}^{p} \alpha_i X_{t-i} + \sum_{j=1}^{m} \beta_j S_{jt} + \varepsilon_t, \tag{1}$$

In this equation, $X_t$ is the demand at time $t$, $S_{jt}$ is the demand uplift state variable at time $t$ taking value of one if demand uplift is in state $j$ at time $t$, and zero otherwise, $\varepsilon_t$ represents a Gaussian White Noise process, $\alpha_i$ and $\beta_j$ are unknown parameters that will be estimated by the time series data and the DUS algorithm output, $p$ is the number of past demand values regressed in the model, and $m$ is the total number of demand uplift states prescribed by the DUS algorithm.

The FSE model has three components: (i) the first two terms on the right-hand side which form an autoregressive model of order $p$ to model the underlying time series (i.e., the time series in the absence of systematic events), (ii) the regression over DUS variables $S_{jt}$ to capture the effect of systematic events, and (iii) the Gaussian white noises to represent the error terms. When there is no event at time period $t$, all state variables in that time period are equal to zero and the FSE model is simplified to an autoregressive model of order $p$. The FSE model assumes that the underlying time series is stationary. Consequently, if there exists a trend in the mean of the underlying time series, the demand variables $X_t$ can be replaced with differenced demand variables in an appropriate lag to convert it to a stationary time series. To demonstrate the application and predictive capabilities of this model, we apply it in two empirical case studies in Section 4.

## 4    Model Validation: Empirical Case Studies

We use empirical sales and promotional data obtained from two FMCG companies in Australia to investigate the validity and industry application of the FSE model. Both companies are major players in the food and beverage industry. We consider demand forecasting for one major product at each company, a more perishable product with a shelf life of a few days and a less perishable product with a shelf life of a few months. Due to data confidentiality reasons, we refer to these companies as Company A and Company B, and their products are denoted by $P_A$ and $P_B$, respectively. In both cases, the supply chain consists of a single supplier (i.e., the case company) satisfying consumer demands of multiple retailers across the country. The time series data and promotional schedules are provided in a weekly format.

### 4.1   Case Study 1: Company A

We have access to 100 weeks of sales data for product $P_A$, including the demand time series, baseline statistical forecasts, final forecasts, and promotional mechanics. The baseline statistical forecasts are produced using an exponential smoothing model which seems to be a common industry practice (Hyndman, Koehler, Ord, & Snyder, 2008). The baseline forecast is regularly adjusted by a panel of experienced forecasters to consider the impact of promotions and other contextual information in the final forecast. The forecasting team consists of four experts, with their work experience varying from 2 to 20 years. There is however no specific procedure in place for identifying and modeling different types of promotions or structuring contextual information of any type.

Figure 1 displays the actual (product demand), baseline statistical forecasts and judgmentally adjusted forecasts which are the final forecasts for product $P_A$. Clearly, the judgmentally adjusted final forecasts appear to be noticeably more accurate than the baseline statistical forecasts as adjustments successfully capture demand uplifts triggered by sales promotions. We are interested in examining the accuracy of the adjusted forecasts and compare the results with the accuracy of forecasts produced by the FSE model.

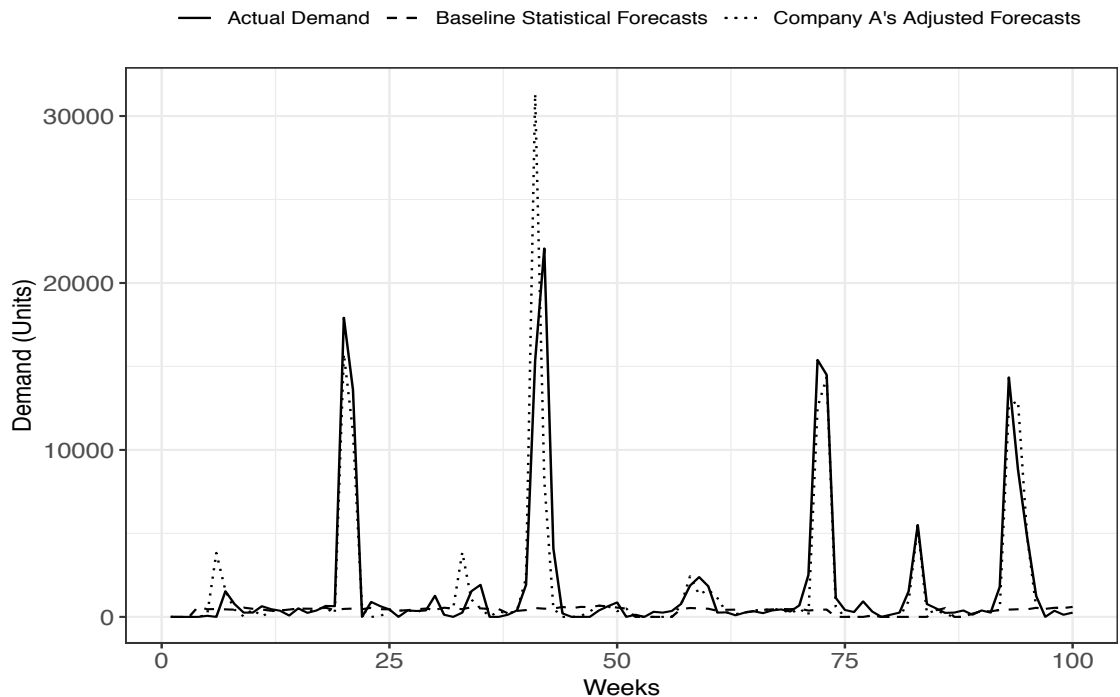

Figure 1: Actual sales and company forecasts for product $P_A$

Different criteria have been used by researchers and practitioners to gauge the effectiveness of a forecasting model. It is of utmost importance to select appropriate error measures when evaluating forecast accuracy (Davydenko & Fildes, 2013). Since the scale of demand dramatically changes and differs for the investigated products, we provide both scale dependent and scale independent metrics to explore the performance of models To ensure consistency and comparability with previous studies (e.g., Baecke et al., 2017; Fildes et al., 2009; Kourentzes & Petropoulos, 2016), we utilize Mean Scaled Absolute Error (MSAE), Mean Absolute Error (MAE), and Mean Absolute Percentage Error (MAPE) as primary accuracy measures in our analyses. Unlike MAE and MAPE which are sensitive to dramatic changes in the scale of data, MSAE is scale independent. The MAE and MAPE are scale dependent and computed using equations (2) and (3), respectively. MSAE is obtained from Equation (4).

$$MAE = \frac{\sum_{t=1}^{n} |f_t - x_t|}{n} \tag{2}$$

$$MAPE = \frac{\sum_{t=1}^{n} \frac{|f_t - x_t|}{x_t}}{n} \tag{3}$$

$$MSAE = \frac{\sum_{t=1}^{n} \frac{|f_t - x_t|}{Total\ demand}}{n} \tag{4}$$

In these equations, $f_t$ is the forecast at time $t$, and $x_t$ is the actual demand at time $t$.

Figure 1 shows dramatic changes in the demand scale for product $P_A$. Descriptive statistics are analyzed separately for promotional and non-promotional periods, and hence the scale of changes has no significant impact on the analysis. Using final adjusted forecasts as the benchmark, Table 1 provides the descriptive statistics for the adjusted forecasts of product $P_A$.

Table 1: Descriptive statistics for product $P_A$

| | |
|---|---|
| Total number of observation weeks | 100 |
| Mean of demand in promotional periods | 9798 |
| Mean of demand in non-promotional periods | 450 |
| MAE for promotional periods | 2348 |
| MAE for non-promotional periods | 334 |
| MAPE for promotional periods | 42% |
| MAPE for non-promotional periods | 16% |

Forecasting experts at Company A stated 'promotion type' and 'display type' as the most influencing factors in promotions. The promotion type can take on one of the two possible levels: 'major' and 'minor'. Promotion type relates to the depth of the discount offered. Although promotions only occur 16 times over the 100 observations, the majority of sales across all observations are realized in these 16 promotional periods (including 8 major and 8 minor promotions). Major and minor promotions are advertised in retailers' weekly catalogues and are typically associated with discounts of approximately 50% and 30% off regular price, respectively. The 'display type' factor can also take one of the four possible levels including: 'entrance', 'Front Gondola End (FGE)', 'other gondola', or 'fixture'. Display type relates to the location in the store where products are displayed. Entrance display types are special out of aisle product displays positioned near the entrance of retail stores. Entrance displays are the most expensive in-store location for product placement but have the most impact on customer purchasing behavior. The FGE displays are at the end of aisles located nearest the front of the store, whereas other gondola displays may be at the end of an aisle but located near the back of the store. Fixture displays are in the aisle of the store that the product is normally located, where there is no additional shelf space for the promotional displays.

After running the DUS algorithm, both promotional mechanics prove to be statistically significant at the 5% level. Although, theoretically, there are eight possible combinations of the levels of the two promotional types, based on available data, the DUS algorithm prescribes five demand uplift states, as shown in Table 2. Those empty cells in Table 2 are corresponding to infeasible combinations in practice.

Table 2: Demand uplift states (average demands uplift in parentheses)

| Promotion Type | Display Type | | | |
|---|---|---|---|---|
| | Entrance | FGE | Other Gondola | Fixture |
| Major | State 1 (19816) | State 2 (14833) | - | - |
| Minor | - | State 3 (5091) | State 4 (3466) | State 5 (4121) |

We use the first 80 weeks of historical data as a training-set, and the last 20 weeks as a test-set to evaluate the performance of the FSE model. More precisely, we fit the FSE model to only include the first 80 weeks of the demand data for product $P_A$ and then use this fitted model to forecast demand for the next 20 time periods. After running the KPSS test to check the stationarity of the underlying demand time series, a p-value of 0.49 was achieved, indicating that at the significance level of 5%, the time series is stationary. By this analysis, an autoregressive model of order two was chosen as the best fit, yielding the lowest AICc. Thus, we set the parameter p=2 in the FSE model. The model was coded and implemented in the R 3.5.0 programming language where parameters are estimated and forecasts for the next 20 time-periods were produced. The estimated parameters in the fitted model were all statistically significant at the 5% level. To demonstrate the validity of the fitted FSE model, diagnostic tests were performed. At the significance level of 5%, the normality hypothesis of residuals failed to be rejected (with a p-value of 8%) and the Ljung-Box test showed that the fitted model shows no lack of fit (with a p-value of 42%).

Figures 2 and 3, respectively, show the absolute and relative errors of Company A's judgmentally adjusted forecasts and the forecasts generated by the fitted FSE model in comparison with actual sales. The two forecasts are compared to actual in the test-set period as shown in Figure 4. The absolute errors and relative errors are calculated from equations (4) and (5), respectively.

$$AE_t = |f_t - x_t| \tag{4}$$

$$RE_t = \frac{f_t - x_t}{x_t} \tag{5}$$

Where $f_t$ is the forecast at time $t$, and $x_t$ is the actual demand at time $t$.

As illustrated in Figures 2-4 and the summary results reported in Table , the FSE model significantly improves the forecast accuracy when compared to Company A's final adjusted forecasts. Table indicates remarkable improvement using different measures such as MSAE (38% improvement), MAE (47% improvement), and MAPE (11% improvement).

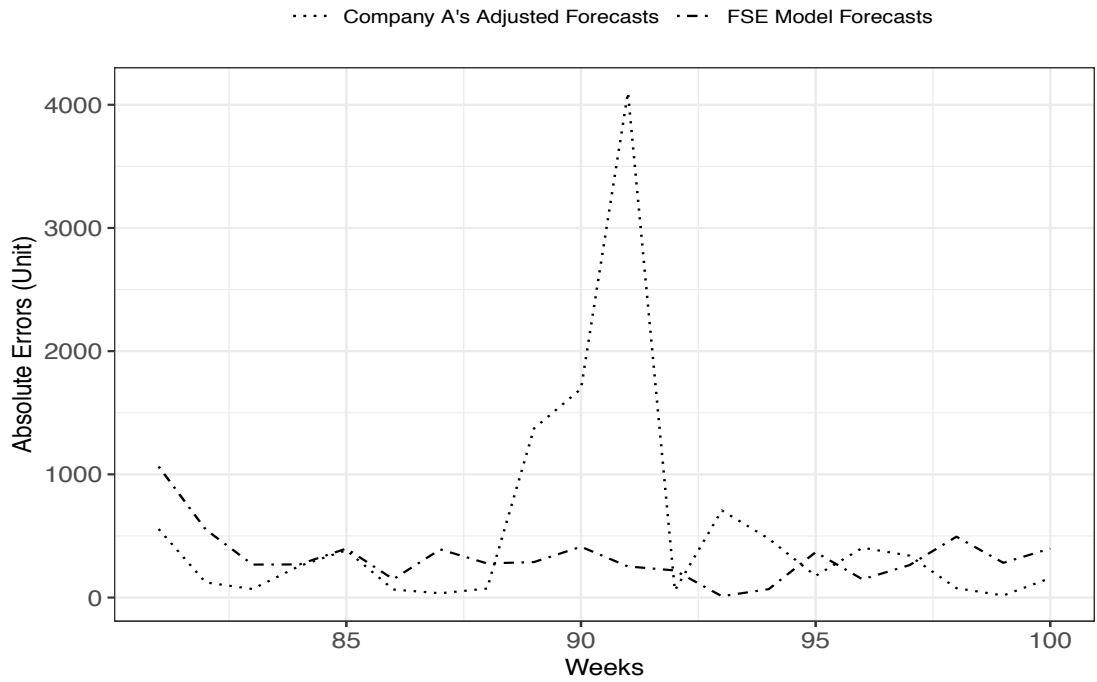

Figure 2: Absolute errors of forecasts compared to actual sales

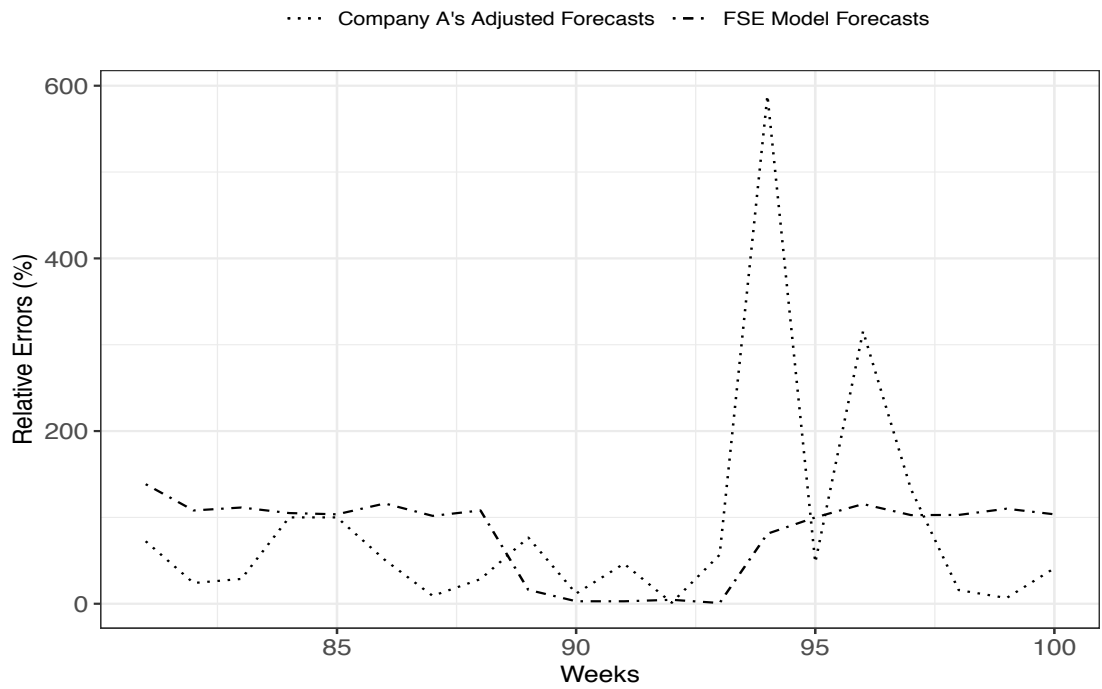

Figure 3: Relative errors of forecasts compared to actual sales

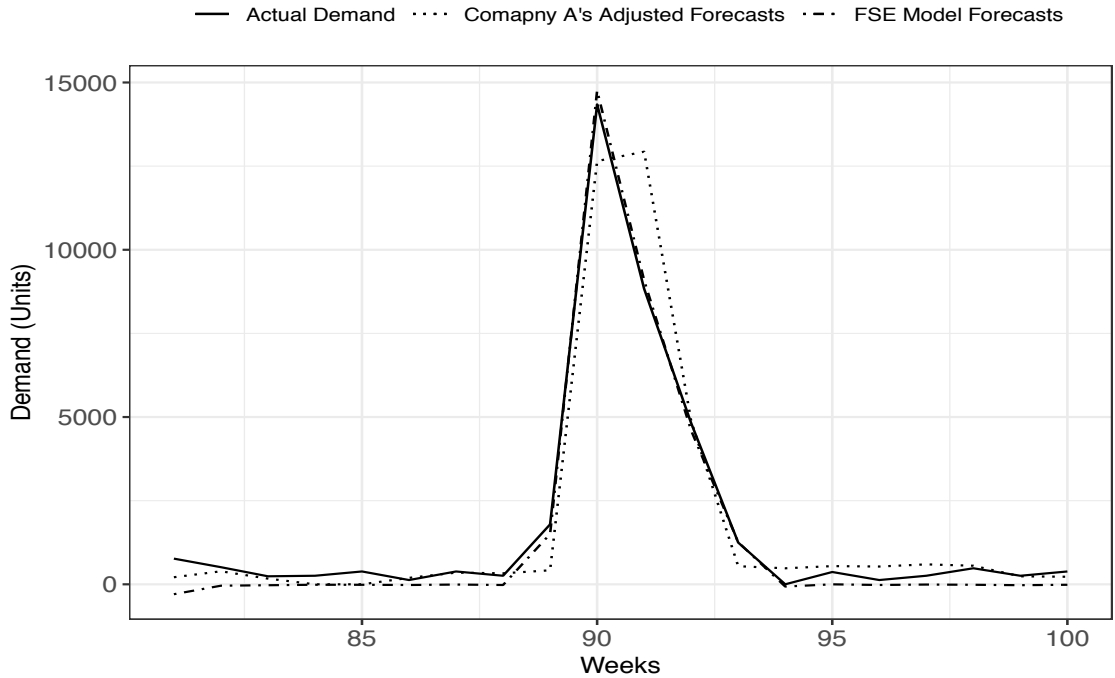

Figure 4: Comparison of the two forecasts and actual sales in the test-set period

Table 3: Forecasting accuracy improvement in the test-set period

| Measure of Error | Company A's Adjusted Forecasts | The FSE Model Forecasts | Improvement in Accuracy |
|---|---|---|---|
| MSAE | 0.18 | 0.11 | 38% |
| MAE | 622.25 | 328.54 | 47% |
| MAPE | 47.11 | 41.70 | 11% |

## 4.2 Case Study 2: Company B

For Company B, we have access to weekly sales data of Product $P_B$ for 120 periods, including the actual demand, baseline statistical forecasts, final forecasts, and promotional mechanics. Analogous to Company A, the baseline statistical forecasting model that is used by company forecasters is a simple exponential smoothing. However, the forecasts obtained from the statistical model are regularly adjusted by the forecasters to account for the effects of sales promotions.

Figure 5: Actual demands and company forecasts for product PFigure 5 shows the actual sales for product PB over 120 weeks as well as Company B's baseline statistical and final adjusted forecasts. Similar to Company A, the adjusted forecasts appear to be noticeably more accurate than the baseline statistical forecasts since baseline forecasts do not consider the substantial uplift in demand caused by sales promotions.

Table   provides the descriptive statistics for $P_B$, using Company B's adjusted forecasts as the benchmark.

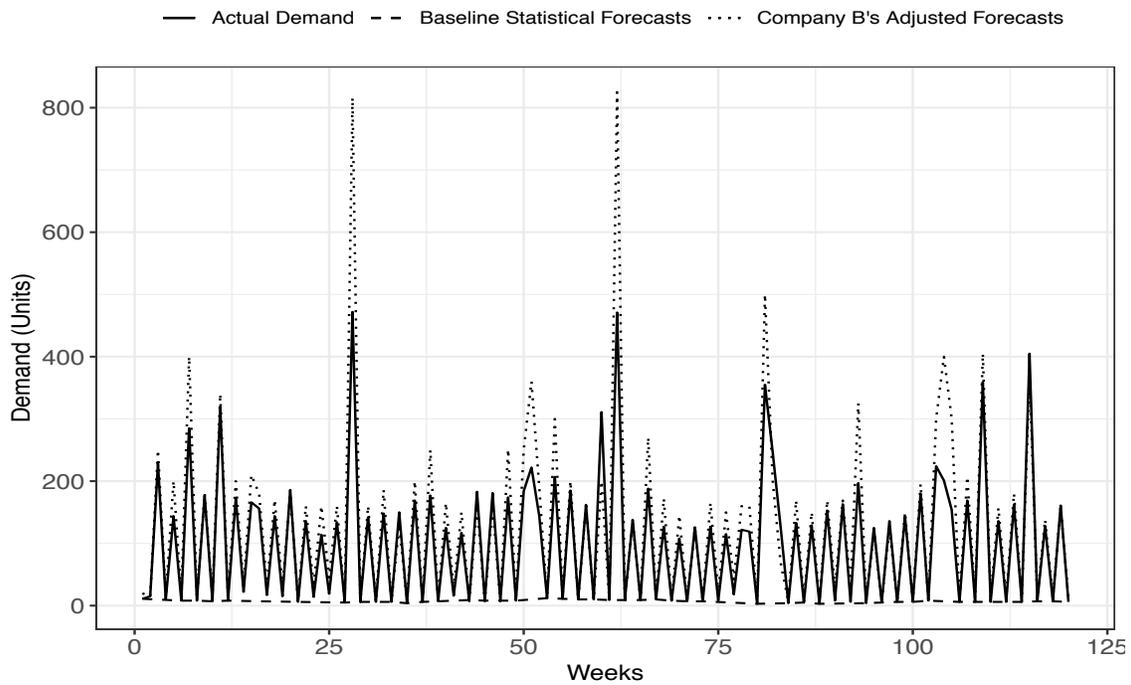

Figure 5: Actual demands and company forecasts for product $P_B$

Table 4: Descriptive statistics for product $P_B$

| | |
|---|---|
| Total number of observation weeks | 120 |
| Mean of demand in promotional periods | 162 |
| Mean of demand in non-promotional periods | 7 |
| MAE for promotional periods | 47 |
| MAE for non-promotional periods | 2.8 |
| MAPE for promotional periods | 41% |
| MAPE for non-promotional periods | 44% |

Forecasting experts at Company B provided the 'promotion type' and 'advertisement type' for product $P_B$. The 'promotion type' factor can take on one of the two possible levels including 'single buy' and 'multiple buy'. There are 60 promotional periods that occur over the 100 observations. 32 of the promotions are 'multiple buy' and 28 are 'single buy' with various discounts. Promotion type relates to whether a promotional discount is offered for a single product or multiple product purchases (e.g., buy one for $25, two for $40, or three for $50). 'Advertisement type' relates to how/where the promotion is advertised, for which the variable can take on one of the three possible levels: 'catalogue', 'minor catalogue' or 'in-store'.

After running the DUS algorithm, both promotional mechanics prove to be statistically significant at the 5% level. Although there are six possible combinations of the levels of the two promotion types, based on available data, the DUS algorithm prescribes five demand uplift states, as shown in Table 5. According to the DUS algorithm, two different combinations including the 'single buy' promotion type with the 'in-store' advertisement type as well as the 'multiple buy' promotion type with the 'in-store' advertisement type are grouped in State 4.

Table 5: Demand uplift states (average uplift values in parentheses)

| Promotion Type | Advertisement Type | | |
|---|---|---|---|
| | Catalogue | In-Store | Minor Catalogue |
| **Single buy** | State 1 (311) | State 4 (16) | State 2 (213.3) |
| **Multiple buy** | State 3 (160.48) | State 5 (15.8) | State 6 (7.3) |

Similar to the process for product $P_A$, the first 100 weeks of historical data is used as a training-set to evaluate the performance of the FSE model in forecasting demand in the last 20 weeks. The KPSS test finds a p-value of 0.77 indicating that the time series is stationary at the significance level of 5%.

An autoregressive model of order two was chosen as the best fit, yielding the lowest AICc. Thus, we set the parameter p=2 in the FSE model. Next, the FSE model was fitted by using the same R-code prepared and implemented for Company A. The estimated parameters in the fitted model were all statistically significant at the 5% level. The diagnostic tests find that the normality hypothesis of residuals failed to be rejected at the significance level of 5% (with a p-value of 28%) and the Ljung-Box test showed that the fitted model does not exhibit lack of fit (with a p-value of 62%).

Figures 6 and 7, respectively, show the absolute and relative errors of Company B's adjusted forecasts and the forecasts generated by the fitted FSE model in comparison with the actual sales. The two forecasts are compared to actual sales in the test-set period as shown in Figure 8. As illustrated in Figures 6-8 and the summary results reported in Table 6, the FSE model significantly improves the forecast accuracy when compared to Company B's judgmentally adjusted forecasts. Table 6 clearly indicates the outstanding improvements using different measures such as MSAE (59% improvement), MAE (55% improvement), and MAPE (14% improvement).

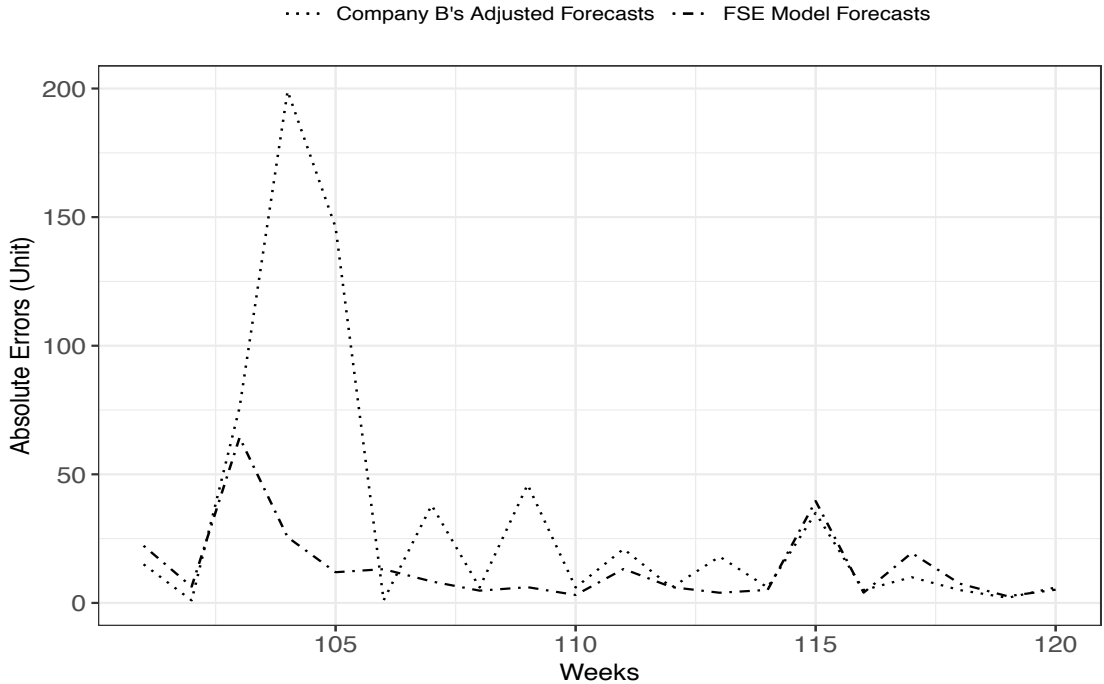

Figure 6: Absolute errors of forecasts compared to actual sales

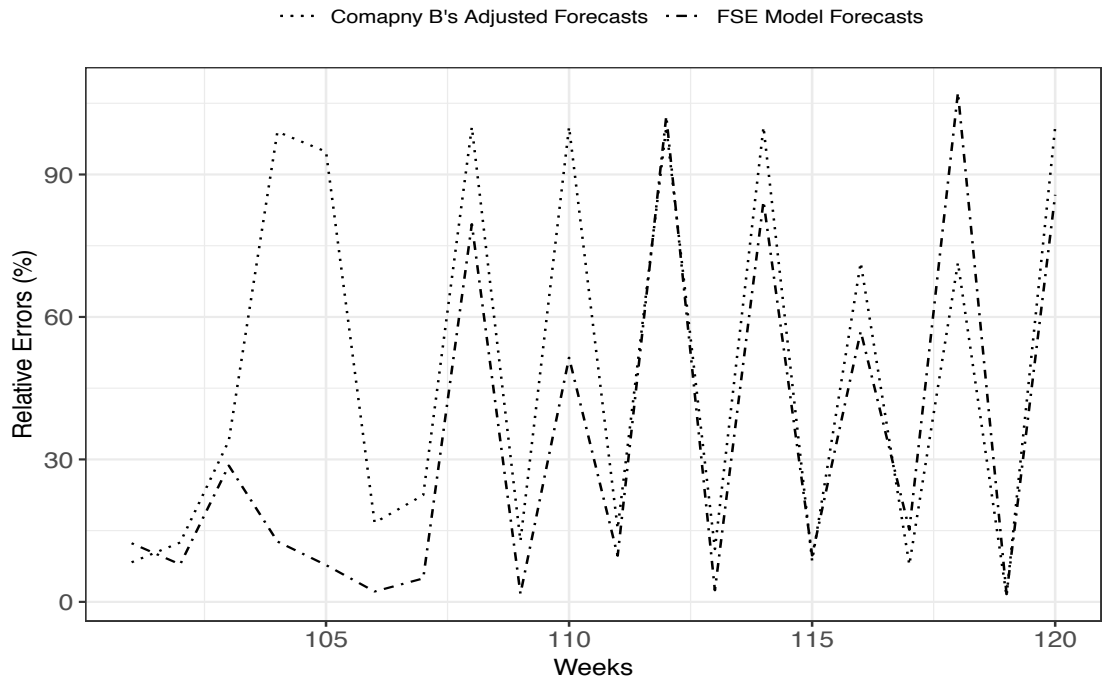

Figure 7: Relative errors of forecasts compared to actual sales

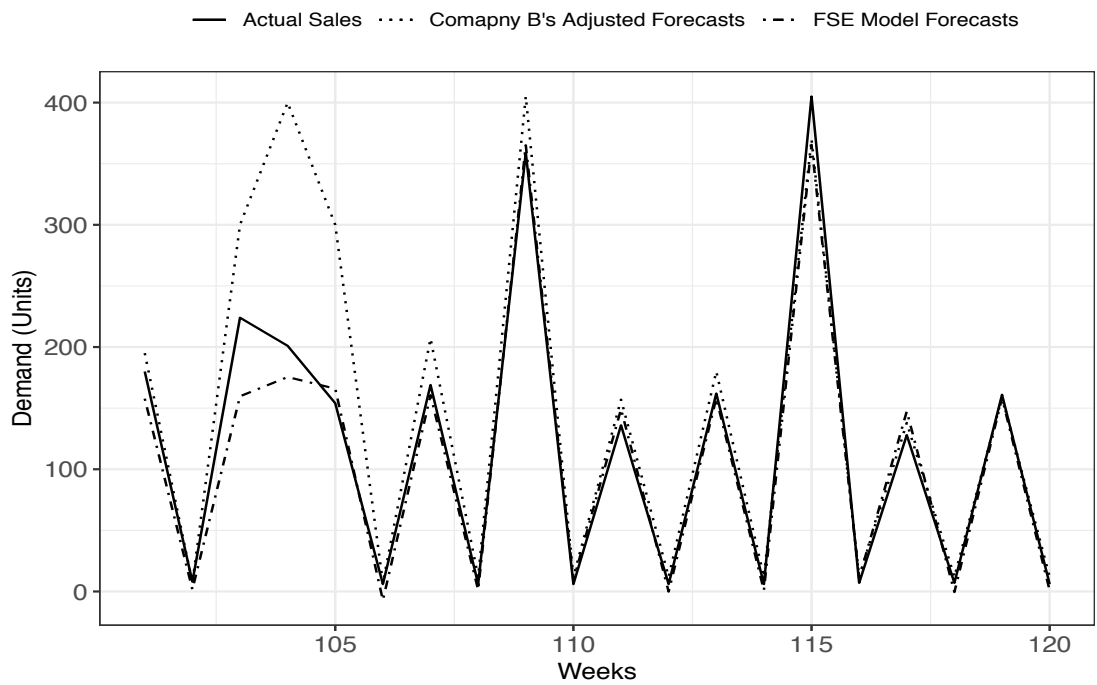

Figure 8: Comparison of the two forecasts and actual sales in the test-set period

Table 6: Forecasting accuracy improvement in the test-set period

| Measure of Error | Company B's Adjusted Forecasts | The FSE Model Forecasts | Improvement in Accuracy |
|---|---|---|---|
| MSAE | 0.32 | 0.13 | 59% |
| MAE | 30.88 | 13.62 | 55% |
| MAPE | 48.60 | 41.65 | 14% |

# 5    Conclusions

In this paper, we propose a time series regression model that structures and embeds systematic contextual information that would otherwise be incorporated with unaided human judgment. Structuring is achieved using an approach, called DUS algorithm, that systematically defines demand uplift states for combinations of levels of factors that contribute to demand uplifts. The factors to consider in DSU algorithm are provided by expert forecasters allowing the model to also benefit from expert knowledge. Once the demand uplift states are identified, they are incorporated into a forecasting model, called FSE, which considers the impact of systematic events to forecast the future demand.

The proposed model and methodology was applied to prepare forecasts in two empirical case studies. We find that the systematic event structuring and forecasting approach can remarkably improve accuracy when compared to the judgmentally made forecasts by the company forecasters (i.e., experienced forecasters making judgmental adjustments to baseline statistical forecasts). Forecast accuracy improvements were demonstrated in both case studies using different measures: MSAE, MAE, and MAPE. We consider this as the key practical contribution of this model since the improved forecast accuracy brings about substantial cost savings according to our industry partners. Indeed, for FMCG companies who forecast for thousands of products on a routine basis, more accurate forecasts – through minimizing judgmental adjustments to incorporate systematic events – can save substantial time and money.

One advantage of the FSE model is the ease with which it allows translating systematic information into demand states. We show in this paper that this can be an effective approach to account for the impact of systematic events such as sales promotions. The difficulty, however, is that the DUS algorithm used to identify demand uplift states is not automated and hence requires substantial time investment when forecasting for a large number of products. We see automating this process – for example, using feedback systems and machine learning approaches – as one important direction for future research. The positive side is that the DUS algorithm that informs the forecasting model does not need to be executed over and over in every forecasting period. Once the demand uplift states are identified, the FSE model can just rerun using the fixed states until new events with different characteristics appear which may then the DUS algorithm to be updated. Another limitation of the proposed approach is the need to have access to sufficient historical data to identify demand uplift states. And for this reason, the FSE model may not be well suited for new products for which the historical sales data is unavailable. This seems to be a common limitation of most statistical models which reply on historical data.

Despite the ease of use, the FSE model can improve forecast accuracy through mitigating the use of unstructured information which seems to be an undeniable part of judgmental forecast adjustments in practice. Part of the uncertainties in demand forecasting roots in how different forecasters evaluate the potential effects of contextual information. For example, different forecasters may have different, potentially contrasting, opinions about the impacts of various promotions. Variations could be caused by psychological reasons or could be due to different experiences, market and supply chain knowledge, access to information, or targets to achieve (Oliva & Watson, 2009). Our model helps overcome such issues as it allows a forecaster to objectively capture the corresponding effects of different quantifiable systematic events such as promotions.

The model and approach developed in this paper is the first attempt to apply concept of regime-switching to define demand states to capture the effects of systematic events. The resulting forecast could be obviously further adjusted by the forecasters to incorporate less quantifiable contextual information and the impact of events that cannot be systematically formulated. Our study thus contributes to the call for further research in structuring the use of human judgment in forecasting (Green & Armstrong, 2007), particularly for products that are prone to sporadic perturbations (De Baets & Harvey, 2018).

Our research also provides insights for innovative FSS design, especially the need for structured support to assist with filtering and integrating information (Fildes et al., 2018). Future research may investigate how this model could be embedded in an FSS so that forecasts for a large number of products are produced semi-automatically. By doing so, judgmental adjustments to forecasts can be more systematic and hassle-free (the use of an FSS can help reduce the cognitive burden on forecasters), and the chance of accounting for the same information twice in demand planning and S&OP would be reduced.

## Acknowledgement

This study was funded by the Australian Research Council (Grant ID: IC140100032). The authors are grateful to Prof Enno Siemsen from Wisconsin School of Business, Prof Marcus O'Connor from The University of Sydney and Prof Elliot Bendoly from The Ohio State University for their invaluable feedback and constructive comments throughout this research.

## References

Adya, M., & Lusk, E. J. (2016). Development and validation of a rule-based time series complexity scoring technique to support design of adaptive forecasting DSS. *Decision Support Systems, 83*, 70-82.

Ali, Ö. G., Sayın, S., Van Woensel, T., & Fransoo, J. (2009). SKU demand forecasting in the presence of promotions. *Expert Systems with Applications, 36*(10), 12340-12348.

Alvarado-Valencia, J., Barrero, L. H., Önkal, D., & Dennerlein, J. T. (2017). Expertise, credibility of system forecasts and integration methods in judgmental demand forecasting. *International Journal of Forecasting, 33*(1), 298-313.

Armstrong, J. S. (1986). The ombudsman: research on forecasting: A Quarter-Century Review, 1960–1984. *Interfaces, 16*(1), 89-109.

Baecke, P., De Baets, S., & Vanderheyden, K. (2017). Investigating the added value of integrating human judgement into statistical demand forecasting systems. *International Journal of Production Economics, 191*, 85-96.

Blattberg, R. C., & Hoch, S. J. (1990). Database models and managerial intuition: 50% model+ 50% manager. *Management Science, 36*(8), 887-899.

Blattberg, R. C., & Neslin, S. A. (1989). Sales promotion: The long and the short of it. *Marketing letters, 1*(1), 81-97.

Cooper, L. G., Baron, P., Levy, W., Swisher, M., & Gogos, P. (1999). PromoCast™: A new forecasting method for promotion planning. *Marketing Science, 18*(3), 301-316.

Davydenko, A., & Fildes, R. (2013). Measuring forecasting accuracy: The case of judgmental adjustments to SKU-level demand forecasts. *International Journal of Forecasting, 29*(3), 510-522.

De Baets, S., & Harvey, N. (2018). Forecasting from time series subject to sporadic perturbations: Effectiveness of different types of forecasting support. *International Journal of Forecasting, 34*(2), 163-180.

Divakar, S., Ratchford, B. T., & Shankar, V. (2005). CHAN4CAST: A Multichannel, Multiregion Sales Forecasting Model and Decision Support System for Consumer Packaged Goods. *Marketing Science, 24*(3), 334-350.

Edmundson, B., Lawrence, M., & O'Connor, M. (1988). The use of non-time series information in sales forecasting: A case study. *Journal of Forecasting, 7*(3), 201-211.

Fildes, R. (1992). The evaluation of extrapolative forecasting methods. *International Journal of Forecasting, 8*(1), 81-98.

Fildes, R., & Goodwin, P. (2007). Against your better judgment? How organizations can improve their use of management judgment in forecasting. *Interfaces, 37*(6), 570-576.

Fildes, R., Goodwin, P., & Lawrence, M. (2006). The design features of forecasting support systems and their effectiveness. *Decision Support Systems, 42*(1), 351-361.

Fildes, R., Goodwin, P., Lawrence, M., & Nikolopoulos, K. (2009). Effective forecasting and judgmental adjustments: an empirical evaluation and strategies for improvement in supply-chain planning. *International Journal of Forecasting, 25*(1), 3-23.

Fildes, R., Goodwin, P., & Önkal, D. (2018). Use and misuse of information in supply chain forecasting of promotion effects. *International Journal of Forecasting.* doi:https://doi.org/10.1016/j.ijforecast.2017.12.006

Fildes, R., Nikolopoulos, K., Crone, S. F., & Syntetos, A. (2008). Forecasting and operational research: a review. *Journal of the Operational Research Society, 59*(9), 1150-1172.

Fischer, I., & Harvey, N. (1999). Combining forecasts: What information do judges need to outperform the simple average? *International Journal of Forecasting, 15*(3), 227-246.

Franses, P. H. (2008). Merging models and experts. *International Journal of Forecasting, 24*(1), 31-33.

Franses, P. H., & Legerstee, R. (2013). Do statistical forecasting models for SKU-level data benefit from including past expert knowledge? *International Journal of Forecasting, 29*(1), 80-87.

Gardner, E. S. (1985). Exponential smoothing: The state of the art. *Journal of Forecasting, 4*(1), 1-28.

Giordani, P., & Kohn, R. (2008). Efficient Bayesian inference for multiple change-point and mixture innovation models. *Journal of Business & Economic Statistics, 26*(1), 66-77.

Giordani, P., Kohn, R., & van Dijk, D. (2007). A unified approach to nonlinearity, structural change, and outliers. *Journal of Econometrics, 137*(1), 112-133.

Goodwin, P. (2002). Integrating management judgment and statistical methods to improve short-term forecasts. *Omega, 30*(2), 127-135.

Goodwin, P., Fildes, R., Lawrence, M., & Stephens, G. (2011). Restrictiveness and guidance in support systems. *Omega, 39*(3), 242-253.

Green, K. C., & Armstrong, J. S. (2007). Structured analogies for forecasting. *International Journal of Forecasting, 23*(3), 365-376.

Green, K. C., & Armstrong, J. S. (2015). Simple versus complex forecasting: The evidence. *Journal of Business Research, 68*(8), 1678-1685.

Hamilton, J. D. (1989). A new approach to the economic analysis of nonstationary time series and the business cycle. *Econometrica: Journal of the Econometric Society,* 357-384.

Hamilton, J. D. (1990). Analysis of time series subject to changes in regime. *Journal of Econometrics, 45*(1-2), 39-70.

Hogarth, R. M., & Makridakis, S. (1981). Forecasting and planning: An evaluation. *Management Science, 27*(2), 115-138.

Huang, T., Fildes, R., & Soopramanien, D. (2014). The value of competitive information in forecasting FMCG retail product sales and the variable selection problem. *European Journal of Operational Research, 237*(2), 738-748.

Hughes, M. (2001). Forecasting practice: organisational issues. *Journal of the Operational Research Society, 52*(2), 143-149.

Hyndman, R. J., & Athanasopoulos, G. (2014). *Forecasting: principles and practice*: OTexts.

Hyndman, R. J., & Koehler, A. B. (2006). Another look at measures of forecast accuracy. *International Journal of Forecasting, 22*(4), 679-688.

Hyndman, R. J., Koehler, A. B., Ord, J. K., & Snyder, R. D. (2008). *Forecasting with exponential smoothing: the state space approach*: Springer Science & Business Media.

Kim, C.-J., Piger, J., & Startz, R. (2008). Estimation of Markov regime-switching regression models with endogenous switching. *Journal of Econometrics, 143*(2), 263-273.

Kourentzes, N., & Petropoulos, F. (2016). Forecasting with multivariate temporal aggregation: The case of promotional modelling. *International Journal of Production Economics, 181*, 145-153.

Kremer, M., Siemsen, E., & Thomas, D. J. (2015). The sum and its parts: Judgmental hierarchical forecasting. *Management Science, 62*(9), 2745-2764.

Lawrence, M., Goodwin, P., O'Connor, M., & Önkal, D. (2006). Judgmental forecasting: A review of progress over the last 25 years. *International Journal of Forecasting, 22*(3), 493-518.

Lawrence, M., O'Connor, M., & Edmundson, B. (2000). A field study of sales forecasting accuracy and processes. *European Journal of Operational Research, 122*(1), 151-160.

Makridakis, S., Andersen, A., Carbone, R., Fildes, R., Hibon, M., Lewandowski, R., . . . Winkler, R. (1982). The accuracy of extrapolation (time series) methods: Results of a forecasting competition. *Journal of Forecasting, 1*(2), 111-153.

Makridakis, S., Chatfield, C., Hibon, M., Lawrence, M., Mills, T., Ord, K., & Simmons, L. F. (1993). The M2-competition: A real-time judgmentally based forecasting study. *International Journal of Forecasting, 9*(1), 5-22.

Makridakis, S., & Hibon, M. (2000). The M3-Competition: results, conclusions and implications. *International Journal of Forecasting, 16*(4), 451-476.

Marmier, F., & Cheikhrouhou, N. (2010). Structuring and integrating human knowledge in demand forecasting: a judgemental adjustment approach. *Production Planning and Control, 21*(4), 399-412.

Moritz, B., Siemsen, E., & Kremer, M. (2014). Judgmental forecasting: Cognitive reflection and decision speed. *Production and Operations Management, 23*(7), 1146-1160.

Murthy, S. K. (1998). Automatic construction of decision trees from data: A multi-disciplinary survey. *Data mining and knowledge discovery, 2*(4), 345-389.

Nikolopoulos, K., Litsa, A., Petropoulos, F., Bougioukos, V., & Khammash, M. (2015). Relative performance of methods for forecasting special events. *Journal of Business Research, 68*(8), 1785-1791.

Nikolopoulos, K., Syntetos, A. A., Boylan, J. E., Petropoulos, F., & Assimakopoulos, V. (2011). An aggregate–disaggregate intermittent demand approach (ADIDA) to forecasting: an empirical proposition and analysis. *Journal of the Operational Research Society, 62*(3), 544-554.

Olafsson, S., Li, X., & Wu, S. (2008). Operations research and data mining. *European Journal of Operational Research, 187*(3), 1429-1448.

Oliva, R., & Watson, N. (2009). Managing functional biases in organizational forecasts: A case study of consensus forecasting in supply chain planning. *Production and Operations Management, 18*(2), 138-151.

Ramanathan, U. (2012). Supply chain collaboration for improved forecast accuracy of promotional sales. *International Journal of Operations & Production Management, 32*(6), 676-695.

Ramanathan, U., & Muyldermans, L. (2010). Identifying demand factors for promotional planning and forecasting: A case of a soft drink company in the UK. *International Journal of Production Economics, 128*(2), 538-545.

Ramanathan, U., & Muyldermans, L. (2011). Identifying the underlying structure of demand during promotions: A structural equation modelling approach. *Expert Systems with Applications, 38*(5), 5544-5552.

Sanders, N. R. (1992). Accuracy of judgmental forecasts: A comparison. *Omega, 20*(3), 353-364.

Sanders, N. R., & Manrodt, K. B. (2003a). The efficacy of using judgmental versus quantitative forecasting methods in practice. *Omega, 31*(6), 511-522.

Sanders, N. R., & Manrodt, K. B. (2003b). Forecasting software in practice: Use, satisfaction, and performance. *Interfaces, 33*(5), 90-93.

Seifert, M., Siemsen, E., Hadida, A. L., & Eisingerich, A. B. (2015). Effective judgmental forecasting in the context of fashion products. *Journal of Operations Management, 36*, 33-45.

Smola, A. J., & Schölkopf, B. (2004). A tutorial on support vector regression. *Statistics and computing, 14*(3), 199-222.

Taylor, J. W. (2003). Exponential smoothing with a damped multiplicative trend. *International Journal of Forecasting, 19*(4), 715-725.

Trapero, J. R., Kourentzes, N., & Fildes, R. (2015). On the identification of sales forecasting models in the presence of promotions. *Journal of the Operational Research Society, 66*(2), 299-307.

Trapero, J. R., Pedregal, D. J., Fildes, R., & Kourentzes, N. (2013). Analysis of judgmental adjustments in the presence of promotions. *International Journal of Forecasting, 29*(2), 234-243.

Van Heerde, H. J., Leeflang, P. S., & Wittink, D. R. (2002). How promotions work: SCAN* PRO-based evolutionary model building. *Schmalenbach Business Review, 54*(3), 198-220.

Webby, R., & O'Connor, M. (1996). Judgemental and statistical time series forecasting: a review of the literature. *International Journal of Forecasting, 12*(1), 91-118.

Zhang, G., Patuwo, B. E., & Hu, M. Y. (1998). Forecasting with artificial neural networks:: The state of the art. *International Journal of Forecasting, 14*(1), 35-62.